\documentclass[preprint,prd,nofootinbib,tightenlines,amsmath,amssymb]{revtex4}
\usepackage{enumerate}
\usepackage{multirow}
\usepackage[utf8]{inputenc}
\usepackage{anysize}
\usepackage{textcomp}
\usepackage{epsfig}
\usepackage{graphics,color} 
\usepackage{verbatim}
\usepackage{afterpage}
\usepackage{amsmath}    
\usepackage{soul}
\usepackage{xcolor,cancel}
\usepackage{geometry}                		
\usepackage[parfill]{parskip}    		
\usepackage{graphicx}				
\usepackage{amssymb}
\usepackage{bm}
\usepackage{todonotes}

\DeclareUnicodeCharacter{00A0}{ }

\usepackage{blindtext}

\catcode`\@=11
\def\sla@#1#2#3#4#5{{%
 \setbox\z@\hbox{$\m@th#4#5$}%
 \setbox\tw@\hbox{$\m@th#4#1$}%
 \dimen4\wd\ifdim\wd\z@<\wd\tw@\tw@\else\z@\fi
 \dimen@\ht\tw@
 \advance\dimen@-\dp\tw@ \advance\dimen@-\ht\z@
 \advance\dimen@\dp\z@
 \divide\dimen@\tw@ \advance\dimen@-#3\ht\tw@
 \advance\dimen@-#3\dp\tw@ \dimen@ii#2\wd\z@
 \raise-\dimen@\hbox to\dimen4{%
 \hss\kern\dimen@ii\box\tw@\kern-\dimen@ii\hss}%
 \llap{\hbox to\dimen4{\hss\box\z@\hss}}}}

\def\cpto{\mathrel {\vcenter {\baselineskip 0pt \kern 0pt
    \hbox{$H_{r.f.}$} \kern 0pt \hbox{$\longrightarrow$} }}}
\def\slashed#1{%
 \expandafter\ifx\csname sla@\string#1\endcsname\relax
{\mathpalette{\sla@/00}{#1}}
\fi}
\def\declareslashed#1#2#3#4#5{%
 \expandafter\def\csname sla@\string#5\endcsname{%
#1{\mathpalette{\sla@{#2}{#3}{#4}}{#5}}}}
 \catcode`\@=12
\declareslashed{}{/}{.08}{0}{D}
 \declareslashed{}{/}{.1}{0}{A}
 \declareslashed{}{/}{0}{-.05}{k}
 \declareslashed{}{/}{.1}{0}{\partial}
 \declareslashed{}{\not}{-.6}{0}{f}

\def\lsim{\mathrel {\vcenter {\baselineskip 0pt \kern 0pt
    \hbox{$<$} \kern 0pt \hbox{$\sim$} }}}
\def\gsim{\mathrel {\vcenter {\baselineskip 0pt \kern 0pt
    \hbox{$>$} \kern 0pt \hbox{$\sim$} }}}

\newcommand{\bea}{\begin{eqnarray}}
\newcommand{\eea}{\end{eqnarray}}

\begin{document}

\baselineskip=15pt
\preprint{}

\title{Relativistic dipole interaction and the topological nature for induced HMW and AC phases}

\author{Xiao-Gang He$^{1,2,3}$\footnote{Electronic address: hexg@phys.ntu.edu.tw}, Bruce McKellar$^{4}$\footnote{Electronic address: bhjmckellar@mac.com }}
\affiliation{
$^{1}$INPAC, Department of Physics and Astronomy,\\
Shanghai Jiao Tong University, Shanghai 200240, China.\\
$^{2}$Department of Physics, National Taiwan University, Taipei 10617, Taiwan. \\
$^{3}$Physics Division, National Center for Theoretical Sciences, Hsinchu 300, Taiwan.\\
$^{4}$ARC Center of Excellence for Particle Physics at the Terascale (CoEPP),\\
School of Physics, University of Melbourne, Victoria 3000, Australia.}

\date{6 February 2017}

\vskip 1cm
\begin{abstract}


In this work we give, for the first time, the full relativistic Lagrangian density describing the motion of 
induced electric  dipoles in the electric fields which induce the dipole,  and the magnetic fields which generate the HMW topological phase.  We then use this  relativistic Lagrangian density to derive the complete set of  conditions for producing topological phases with induced dipoles.  We also give the relativistic Lagrangian density describing the motion of 
induced magnetic  dipoles in the magnetic  fields which induce the dipole,  and the electric  fields which generate the AC  topological phase, and derive the conditions for this AC phase to be topological.   These conditions have been incompletely  discussed in previous studies. We note that, in both the AC and HMW cases, the topological phases are generated by
the cross product of electric and magnetic fields in the form $\bm{B} \times \bm{E}$ which reinforces the dual nature of these two topological phases.

\end{abstract}

\pacs{PACS numbers: }

\maketitle

The topological phase is one of the most important aspects of quantum mechanics which distinguish quantum from classical mechanics. Some of the key examples come from quantum mechanical electromagnetic interactions. There are three quantum mechanical electromagnetic topological phases, the Aharonov-Bohm (AB) phase\cite{AB}, 
the Aharonov-Casher (AC) Phase\cite{AC} and the He-McKellar-Wilkens (HMW) phase\cite{HM,W, Dowling}, which have been experimentally verified. These phases reside in the phase factor of the wave function. A common feature of the topological  nature of these phases is that when a particle, carrying a  certain ``charge''  which induces interactions with  external electric and/or magnetic fields,  travels through regions  where, in classical sense, no forces act on the particle, but when it encircles a closed path which contains certain field configurations (FC) which the particle does not enter, the wave function develops a non-trivial phase independent of the particular path travelled as long as it encloses the given FC\cite{MHK}. In the AB case, the charge is the electric charge and the FC is magnetic flux. In AC and HWM cases, the charges are magnetic dipole and electric dipole 
and the FCs are field configurations in which the vector product of the magnetic or electric dipole  and the electric  or magnetic field has a  non-vanishing curl. These non-trivial phases cause interference effects which have been observed experimentally, the topological AB phase by  Chambers\cite{Chambers} and by Tonomura \emph{et al.}\cite{Tom86}, the topological AC phase by Cimmino \textit{et al}\cite{PRL63_380} and the Toulouse group\cite{Gillot2014}, and the topological HMW effect by the Toulouse group\cite{Lepoutre2012}, and the Tokyo Atom Interferometry Group\cite{kumiya-2016}.  For the HMW effect, because no atom has an observable electric dipole moment, it is necessary to induce the electric dipole moment by applying an electric field\cite{W}.  Using an induced electric dipole moving in a magnetic field also means that the phase can be topological without needing a magnetic monopole source of the magnetic field\cite{WHW}.  Recent reviews of electromagnetic topological phases have been given by BMcK in ref. \cite{Mckellar:2016zzf} and ref \cite{McKellar:2014loa}.

A crucial ingredient in identifying a topological phase is to analyze the relevant Hamiltonian governing the motion of a particle to see if there are certain configurations so that there is a term (terms) $H_{top}$ which exhibits the feature that they do not exert force on the particle and therefore can be transformed into the wave function producing a phase factor $e^{-iH_{top} \Delta t}$. When the particle is traveling a closed path taking a time T, one integrates the phase factor $\int_0^T H_{top} dt$. One can translate this into an integral along the path, 
$\oint_{\mathcal{C}} \boldsymbol{ T \cdot d r}$.  One then checks to see if this integral has a value independent of the path travelled. In order to have a topological phase certain conditions have to be satisfied. Without a clear understanding of the conditions one may make false interpretations of the observation. 

The Aharonov-Bohm phase is always topological in its nature as long as the path of the charged particle encloses some magnetic flux. For the AC and HMW phases topological phases can develop only under certain circumstances. Essential conditions are that there is translational symmetry in one direction, and that the dipole moment in question is normal to both the direction of motion and the field with which it interacts. For the HMW phase it is particularly important to understand the constraints that must be satisfied by the  electromagnetic fields through which the induced dipole is traveling. In the following we clarify  these conditions which have not been accurately discussed for these induced dipoles.  Previous discussions have been based on non-relativistic Lagrangian for interaction of a dipole with electromagnetic fields.  In this work we give, for the first time, the relativistic Lagrangian density describing the motion of 
induced electric (and magnetic) dipoles in electric (magnetic) fields which induce the dipole, and magnetic (electric) fields which generate the HMW (AC) topological phases.  We then use this  relativistic Lagrangian density to derive the complete set of  conditions for producing topological phases with induced dipoles in the non-relativistic regime.  These conditions have been incompletely  discussed in previous studies.   However they have been met by the experiments which observed the HMW phase.

Because of its practical importance, we will first discuss the HMW phase conditions, and then discuss the case of AC phase.

 The analysis should be based on induced dipoles from the start, which is what Wei, Han and Wei actually did\cite{WHW}.
They based their analysis, as did Wilkens\cite{W},  on the effective electric field seen by a particle moving in an magnetic field --- called the R\"ontgen  field $\mathbf{E_{\mbox{R}} = v \times B}$, which is added to the laboratory electric field $\bm{E}$ to obtain the electric field $\bm{E}_0 = \mathbf{(E + v\times B)}$ experienced by the dipole in its rest frame.   Through out this paper, $c=1$ and $\hbar = 1$ units will be used. If the polarisability of the atom is $\alpha$ its electric dipole moment is 
\begin{equation}
\mathbf{d} = \alpha \mathbf{(E + v\times B)}\;.
\end{equation}
The Lagrangian is then 
\begin{eqnarray}
 \mathcal{L} &=& \frac{1}{2}m \mathbf{v}^2 +\frac{1}{2}\alpha (\mathbf{E + v\times B})^2\nonumber\\
 &=&  \frac{1}{2}(m + \alpha \mathbf{B}^2) \mathbf{v}^2 +\frac{1}{2}\alpha \left (2 \mathbf{v \cdot (B \times E)}+\mathbf{E}^2  -\alpha (\mathbf{v\cdot B})^2\right )\;.
\end{eqnarray}

 Wei, Han, and Wei obtain their form of the HMW phase  by
neglecting the terms $\alpha\mathbf{E}^2$, verifying that $ \alpha (\mathbf{B})^2 \ll m$, so that it may also be neglected, and ensuring that  the experimental configuration is such that $\mathbf{v} \perp \mathbf{B}$, so that $\mathbf{v} \cdot \mathbf{B}$.  We follow their example and  the resulting Schr\"odinger equation is
\begin{equation}
\frac{1}{2m} \left( -i\nabla - \alpha( \mathbf{B \times E})\right)^2 \psi = 0\;,
\end{equation} 
 which can be transformed to the field free Schr\"odinger  equation by a phase transformation with the phase
\begin{equation}
\chi_{_{WHW}} = \alpha \int_{\mathcal{C}} \mathbf{B \times E \cdot dr}\;.
\end{equation}

This is a topological phase if 
$$\mbox{curl}\;(\mathbf{B \times E) = B\;\mbox{div}\,E - E\; \mbox{div}\,B +(B\cdot \nabla) E - (E\cdot \nabla) B}\;,$$
vanishes in the interference region and is non-zero in the excluded region.    Now electric charges, as sources of $\mathbf{E}$, can generate the topological phase through the first term above\cite{W}. Inducing the electric dipole removes the link to magnetic monopoles.
So far we have not specified any condition on the electric field.  Our previous relativistic analysis\cite{HM} suggests that the electric field should also be normal to the velocity.  Why has this not come out in the analysis of Wei, Han, and Wei\cite{WHW}?
 
In an attempt to answer this question, it would seem to be more reliable to obtain the relativistic corrections by taking the low velocity limit of a fully relativistic theory, rather than trying to introduce the corrections into the non-relativistic result.  That is the approach we now adopt.

Minkowski\cite{Mink} is the standard reference for the relativistic treatment of materials.  There are accessible accounts in Pauli\cite{Pauli} and M\o ller\cite{mol}.  Minkowski's proposal is that the relativistic version of $\mathbf{D}$ and $\mathbf{H}$ is the tensor $G_{\mu\nu}$ obtained by replacing $\mathbf{E}$ and $\mathbf{B}$ in $F_{\mu\nu}$ by $\mathbf{D}$ and $\mathbf{H}$.  The Lagrangian density is then $-G_{\mu\nu}F^{\mu\nu}$ (up to some constant factor).

As (in Heaviside-Lorentz units) $$\mathbf{D} =  \mathbf{E} + \mathbf{P}\;,\;\;\mathbf{H} =  \mathbf{B}  - \mathbf{M}\;,$$ we should introduce a tensor (which Becker and Sauter\cite{BS} call the moments tensor) $K_{\mu\nu}$ constructed from $F_{\mu\nu}$ by replacing $\mathbf{E}$ with $\mathbf{P}$ and $\mathbf{B}$ with -$\mathbf{M}$. 
A moments tensor can be constructed from the electric and magnetic polarisation density  of material bodies or the electric and magnetic dipole moments of individual atoms.   We will use $K_{\mu\nu}$ as the moments tensor of atoms.
This clearly shows that in relativity electric and magnetic moments get mixed up.  This is nicely explained (with examples) by Becker and Sauter.

For now we will ignore intrinsic moments which are proportional to the spin of the particle, and  consider only induced moments, which are proportional to the applied fields.  We have to get to the generalisation of $\mathbf{P}  = \alpha \mathbf{E}$,  and $\mathbf{M} = \chi \mathbf{B}$, where $\alpha$ is the electric polarisability and $\chi$ is the magnetic susceptibility, which hold in the rest frame of the material.  Following Minkowski we write, with  $u_\mu$ as the four velocity  of the moving particle,
\begin{equation}
u^{\mu}K_{\mu\nu} = \alpha u^{\mu}F_{\mu\nu} \quad\quad \mbox{and} \quad\quad u^{\mu}\tilde{K}_{\mu\nu} = \chi u^{\mu}\tilde{F}_{\mu\nu},  \label{min}
\end{equation}
  which are identical to the above in the rest frame, and are  tensor equations, so they are the correct generalisation.   

In the rest frame the electric and magnetic fields are the spatial components of 
\begin{equation}
E^\mu = u_\nu F^{\mu\nu} \quad \mbox{and}\quad B^\mu = u_\nu\tilde{F}^{\mu\nu}\;,
\end{equation}
and, in the sense that $qE^\mu$ and $gB^\mu$ are the Minkowski four-force on a test charge $q$ and a test magnetic monopole $g$ in a frame in which they are moving with four-velocity $u^\mu$,  $E^\mu$ and $B^\mu$ are appropriate relativistic generalisations of $\mathbf{E}$ and $\mathbf{B}$ respectively. In particular $qE^\mu$ is simply the Minkowski four-force version of the Lorentz force on the charged particle.  Its spatial component $q\gamma (\boldsymbol{E + v \times B})$ is the Lorentz force on the charge, and its time component $q\gamma \boldsymbol{v \cdot E} $ is the work done on the charged particle. Here $\gamma =1/\sqrt{1-v^2}$.

Similarly 
\begin{equation}
P^\mu = u_\nu K^{\mu\nu} \quad \mbox{and} \quad M^\mu = u_\nu\tilde{K}^{\mu\nu}\;,
\end{equation}
are appropriate relativistic generalisations of $\mathbf{P}$ and $\mathbf{M}$ respectively.

We re-write equation (\ref{min}) as 
\begin{equation}
P^\mu = \alpha E^\mu \quad\quad \mbox{and} \quad\quad M^\mu = \chi B^\mu\;,
\end{equation}  
as the relationship between the induced electric moment and the electric field and the induced magnetic moment and the magnetic field.  In the rest frame they are identical to the non-relativistic equations, and are tensor equations, so they are the correct generalisation, at least in the simplified situation that the polarisability and the susceptibility are scalars.  Pauli's discussion (in \S 34) suggests that this is an adequate approximation, and we will use it.

M\o ller shows how to construct $K_{\mu\nu}$ from $P^\mu$ and $M^\mu$, and hence the electromagnetic field.  The result is
\begin{equation}
K_{\mu\nu} = \alpha\left\{u_\mu E_{\nu}  - u_\nu E_{\mu}\right\} + \chi \epsilon_{\mu\nu\kappa\lambda}B^{\kappa} u^\lambda.
\end{equation}
Consider only the induced electric dipole moment, and set $\chi = 0$.  Then with the electric 4-vector
\begin{eqnarray}
E^\mu  & =  & \gamma\left(\mathbf{E\cdot v},  \mathbf{E + v \times B}\right)\;,\;\;
K_{\mu\nu} =  \alpha \left\{ u_\mu E_\nu - u_\nu E_\mu \right\},
\end{eqnarray}
the interaction Lagrangian is 
\begin{equation}
{\mathcal L}_{\rm{int}} = - \frac{1}{4}K^{\mu\nu}F_{\mu\nu} = -\frac{1}{2}\alpha E_{\mu}E^{\mu} \quad =  \frac{1}{2}\alpha \gamma^2 \left\{(\mathbf{E + v \times B})^2 - (\mathbf{E\cdot v})^2\right\}\;.
\end{equation}

We find that, keeping only those terms of order  $v^2$ which introduce new interactions, the correct non-relativistic reduction of the interaction Lagrangian  of an electrically polarisable moving neutral atom  in electric and magnetic fields is 
\begin{equation}
{\mathcal L}_{\rm{int}} = \frac{1}{2}\alpha \left\{(\mathbf{E + v \times B})^2 - (\mathbf{E\cdot v})^2\right\}\;. \label{ff}
\end{equation}
The term $(\mathbf{E\cdot v})^2$ was omitted by Wei, Han and Wei\cite{WHW}, but, as long as $\mathbf{E\cdot v} = 0$, the phase is still topological. If this is not the case, there are additional terms  which confuse the nature of effects for observation.

We now have the condition that the particle velocity is normal to both of the electric and magnetic fields, re-establishing  the planar geometry.

We emphasise again that, for induced electric dipoles, one can obtain the HMW topological phase without magnetic monopoles.

 Now that we have derived the conditions for an induced electric dipole, we can see how they connect with the  He-McKellar\cite{HM, ALSO} derivation of the HMW phase of an intrinsic electric dipole $\bm{d}$ moving in a magnetic field $\bm{B}$ with a velocity $\bm{v}$, which shows that the phase is given by
 \begin{eqnarray}
\chi_{_{HMW}} & = &  \oint_{\mathcal{C}} \boldsymbol{ T \cdot d r}\;,\;\;
\mbox{with}\; \boldsymbol{ T}  = \boldsymbol{B \times d}\;. \label{hmw-1}
\end{eqnarray}
 In this derivation $\boldsymbol{B \times d}$ is on the plane normal to $\bm{d}$, so the situation has essentially two spatial dimensions.  For a constant $\bm{d}$ 
 \begin{equation}
 \bm{\nabla \times (B \times d) = -\left(d \nabla \cdot B + (d \cdot \nabla)B\right)}.
 \end{equation}
 If we have monopoles as a source of $\bm{B}$ there will be a non-vanishing curl of $\bm{B\times d}$, but it is also possible to have a non-vanishing curl in a suitably shaped magnetic field, as discussed by McKellar, He and Klein\cite{MHK}.
 
 When the electric dipole is induced by an  electric field $\bm{E}$, $\bm{d} = \alpha \bm{E}$, where $\alpha$ is the electric polarisability of the particle.
This suggests that we should set $\boldsymbol{ T}  =  \alpha \boldsymbol{B \times E}$,  which  has the wonderful advantage that 
 \begin{equation}
 \boldsymbol{ \nabla \times (B \times } \boldsymbol{E}]) =  \left(\boldsymbol{B\;\mbox{div}\,E - E\; \mbox{div}\,B +(B\cdot \nabla) E - (E\cdot \nabla) B}\right)\;,
 \end{equation}
 so that 
 $\mbox{div}\bm{E} \ne 0$ in the excluded region can give $\boldsymbol{ \nabla \times (B \times } [\alpha \boldsymbol{E}]) \ne 0$ in that region, and, as was derived \textit{ab initio} above,  we can have a topological phase without the magnetic monopoles  used in the original HWM papers.

The above results can also be obtained by first constructing the Lagrangian density of an electric dipole interaction with electromagnetic fields and then taking the electric dipole interaction to be induced by the electromagnetic fields applied. 

In the rest frame, the Lagrangian density of an electric dipole $\bm{ d}$ is proportional to its spin and therefore $\bm{d^\prime} = d\bm {s^\prime}$ and electric field $\bm{E^\prime}$ is given by
${\mathcal L}_{\rm{int}} = d \bm{s^\prime}\cdot \bm{E^\prime}$. Using the known transformation properties of a four spin vector in the moving frame $s_\alpha = (\gamma \bm{v} \cdot \bm{s^\prime}, \bm{s^\prime} + (\gamma^2/(1+\gamma))(\bm{v}\cdot\bm{s^\prime})\bm{v})$, the Lagrangian density can be written as $du_\mu s_\nu F^{\mu \nu} = - d s_\mu E^\mu$, one obtains 
\begin{eqnarray}
{\mathcal L}_{\rm{int}}  = d \gamma \bm{s} \cdot (\bm{E} + \bm{v} \times \bm{B} - \bm{v} (\bm{v} \cdot \bm{E}))\;. \label{tt}
\end{eqnarray}{
The middle term induces the HMW phase.  Note that in the above we are considering an intrinsic dipole moment of an elementary particle, so $d \ne 0$ implies $P$ and $T$ violation.

To use the above equation for induced electric dipole, one just needs to identify the corresponding induced dipole density operator.
In the rest frame, the induced electric dipole  given by $\bm{d^\prime}$ is  proportional to $\alpha \bm{E^\prime}$, and we emphasise that in this relationship, $P$ and $T$ invariance holds.  The Lagrangian density is then
${\mathcal L}_{\rm{int}} = ({1/2}) \alpha \bm {E^\prime} \cdot \bm{ E^\prime}$. Therefore, in the rest frame the induced electric dipole is identified as $\bm{d^\prime} = \alpha \bm{E^\prime}$. 
Using $\bm{ s} = \bm{s^\prime} + (\gamma^2/(1+\gamma)) (\bm{v} \cdot \bm{s^\prime}) \bm{ v}$ and then expressing $\bm{s^\prime}$ in terms of $\bm{E}$ and $\bm{B}$, one obtains $d \bm{ s} = \alpha \gamma (\bm{E} +\bm{v}\times \bm{B})$. Finally inserting this expression into eq.\ref{tt}, we obtain the interaction Lagrangian density as that given in eq.\ref{ff}. We are relying on the continuous Lorentz transformation properties of the 4-spin in this calculation, and the fact that the 4-spin and the induced 4-dipole have opposite discrete T and P transformation properties does not invalidate the calculation.

This observation provides a complete relativistic derivation without the use of the specific Dirac equation for spin half and Bargmann-Wigner equation for arbitrary spin equations of motion for the AC and HMW topological phases\cite{AC,HM,ALSO} with permanent magnetic and electric dipole moments, respectively.

Replacing $\bm{E}$ by  $\bm{B}$, $\bm{ B}$ by  $- \bm{ E}$ and $\alpha$ by $\chi$, one obtains the the interaction Lagrangian density for an induced magnetic dipole relevant for the AC phase\cite{AC}. Therefore in the induced AC phase case, the condition that the particle velocity is  normal to both of the electric and magnetic fields is also required. The geometry is again  the planar geometry.

Note that in the induced electric dipole and magnetic dipole interaction with electromagnetic fields, the AC and HMW phases are all generated by $\bm{ E} \times \bm{B}$, this new insight  reinforces the dual nature of AC and HMW phases.
\\

\noindent
$\bm{Acknowledgements}$

XGH was supported in part by MOE Academic Excellent Program (Grant No. 105R891505), NCTS and MOST of ROC (Grant No. MOST104-2112-M-002-015-MY3), and in part by NSFC (Grant Nos. 11175115) 
and Shanghai Science and Technology Commission (Grant No. 15DZ2272100) of PRC. B.H.J.M was supported in part by by the Australian Research Council Grant to the ARC Centre of Excellence for Particle Physics at the Terascale (CoEPP). 
BHJMcK thanks NTU and NCTS for hospitality where this work was done.
\\

\noindent
{\bf Note Added}

We thank the referee for bringing the recent paper\cite{kaima} by Kai Ma to our attention, in which AC phases associated with induced magnetic dipoles and HMW phases associated with induced electric dipoles are considered, following the methods of Wei, Han and Wei.  The conditions on orthogonality of $\mathbf{v}$, $\mathbf{E}$ and $\mathbf{B}$ are not discussed, and the fully relativistic approach described here is not used.  The paper makes the interesting suggestion that the discrepancy between the calculated phase and that observed by the Toulouse group could be due to an AC phase from and induced magnetic dipole.


\begin{thebibliography}{999}

\bibitem{AB} Y. Aharonov and D. Bohm, Phys. Rev 115, 485-491 (1959).  W. Ehrenberg and R.E. Siday, Proc. Phys. Soc. B 62, 8- 20(1949).
\bibitem{AC} Y. Aharonov and A. Casher, Phys. Rev. Lett. 53, 319-321 (1984).
\bibitem{HM} X.-G. He and B. H. J. McKellar, Phys. Rev. A 47, 3424 (1993).
\bibitem{W} M. Wilkens, Phys. Rev. Lett. 72, 5 (1994).
\bibitem{Dowling} J. Dowling, C. Williams and J. Franson, Phys. Rev. Lett. 83, 2486(1999).

\bibitem{MHK}
 B. H. J. McKellar, X-G. He and  A. G. Klein,
AIP Conference Proceedings 1588, 59 (2014).

  
\bibitem{Chambers} G. Chambers, Phys. Rev. Lett. 5, 3(1960).

\bibitem{Tom86} A. Tonomura, N. Oskabe, T. Matsuda, T. Kawasaki and J. Endo, Phys. Rev. Lett. 56, 297(1986).

\bibitem{PRL63_380} A. Cimmino, G.I. Opat, A.G. Klein, H. Kaiser, S.A.
Werner, M. Arif and R. Cothier, Phys. Rev. Lett. 63, 380 (1989).

\bibitem{Gillot2014} J. Gillot, S. Lepoutre and A. Gauguet, Eur. Phys. J. D, 68, 168(2014).

\bibitem{Lepoutre2012} S. Lepoutre, A. Gauguet, G. Trnec, M. Buechner and J.
Vigu\'e, Phys. Rev. Lett. 109, 120404 (2012); 
J. Gillot, S. Lepoutre, A. Gauguet, M. Buechner and J. Vigu\'e, Phys. Rev. Lett. 111, 030401(2013);
S. Lepoutre, A. Gauguet, M. Buechner and J.
Vigu\'e, Phys. Rev. A88, 043627 (2013); S. Lepoutre, J. Gillot, A. Gauguet, M. Buechner and J. Vigu\'e, Phys. Rev. A88, 043628. 

\bibitem{kumiya-2016} T. Kumiya, A. Akentyev, Y. Mori, J. Ichimura and A. Morinaga, Phys. Rev. A 93, 023637(2016).

\bibitem{WHW} H. Wei, R. Han, and X. Wei, Phys. Rev. Lett. 75, 2071
(1995).  
 
\bibitem{Mckellar:2016zzf} 
  B.~H.~J.~McKellar,
  Int.\ J.\ Mod.\ Phys.\ A {\bf 31}, 1630004 (2016).
  
\bibitem{McKellar:2014loa} 
  B.~H.~J.~McKellar,
  The Universe {\bf 2}, no. 4, 6 (2014). 
  
\bibitem{ALSO}
X.-G. He and B. H. J. McKellar, Phys. Lett. B 256, 250-254 (1991);
X.-G. He and B. H. J. McKellar, Phys. Rev. A, 64, 022012 (2001);
X.-G. He and B. H. J. McKellar, Phys. Lett. B 559, 263-269 (2003).

\bibitem{Mink} H. Minkowski \textit{G\"ott. Nachr.}, p53 (1908).
\bibitem{Pauli} W. Pauli \textit{Theory of Relativity}, Dover reprint 1981 see \S 33.
\bibitem{mol} C. M\o ller \textit{The Theory of Relativity} Oxford 1957.
\bibitem{BS} R. Becker and F. Sauter, \textit{Electromagnetic Fields and Interactions}, Vol 1 Chapter E III. Blackie 1964.

\bibitem{kaima} Kai Ma, Chin. Phys. Lett. vol 33, 090304 (2016) [arXiv:1608.07862 [quant-ph]].

\end{thebibliography}
\end{document}